# Quantum Annealing Approach for the Optimal Real-time Traffic Control using QUBO

Amit Singh*,†
National Center for High-Performance Computing
National Applied Research Laboratories
Hsinchu 300, Taiwan
amit.en06@nycu.edu.tw

Chun-Yu Lin
National Center for High-Performance Computing
National Applied Research Laboratories
Hsinchu 300, Taiwan
lincy@narlabs.org.tw

Chung-I Huang
National Center for High-Performance Computing
National Applied Research Laboratories
Hsinchu 300, Taiwan
schumi@narlabs.org.tw

Fang-Pang Lin
National Center for High-Performance Computing
National Applied Research Laboratories
Hsinchu 300, Taiwan
fplin@narlabs.org.tw

*Abstract*— **Traffic congestion is one of the major issues in urban areas, particularly when traffic loads exceed the road's capacity, resulting in higher petrol consumption and carbon emissions as well as delays and stress for road users. In Asia, the traffic situation can be further deteriorated by road sharing of scooters. How to control the traffic flow to mitigate the congestion has been one of the central issues in transportation research. In this study, we employ a quantum annealing approach to optimize the traffic signals control at a real-life intersection with mixed traffic flows of vehicles and scooters. Considering traffic flow is a continuous and emerging phenomenon, we used quadratic unconstrained binary optimization (QUBO) formalism for traffic optimization, which has a natural equivalence to the Ising model and can be solved efficiently on the quantum annealers, quantum computers or digital annealers. In this article, we first applied the QUBO traffic optimization to artificially generated traffic for a simple intersection, and then we used real-time traffic data to simulate a real "Dongda-Keyuan" intersection with dedicated cars and scooter lanes, as well as mixed scooter and car lanes. We introduced two types of traffic light control systems for traffic optimization: C-QUBO and QUBO. Our rigorous QUBO optimizations show that C-QUBO and QUBO outperform the commonly used fixed cycle method, with QUBO outperforming C-QUBO in some instances. It has been found that QUBO optimization significantly relieves traffic congestion for the unbalanced traffic volume. Furthermore, we found that dynamic changes in traffic light signal duration greatly reduce traffic congestion.**

***Keywords—real-time optimization, traffic optimization, traffic signal control, quantum annealing, quantum computing, QUBO***

## I. Introduction

Traffic congestions continue to grow worldwide, increasing travel delay, fuel consumption, and pollution [1]. This is particularly important at gatherings that bring together large groups of people for short periods (like athletic activities, concerts, and rush hour) and may create substantial disruptions to cities' transit networks, causing delays for residents. In 2017, traffic congestion caused metropolitan Americans to spend an additional 8.8 billion hours of travel time and purchase an additional 12.5 billion litres of gasoline, which cost around $179 billion [2]. With the transient existence of rush hour, fixed transit infrastructure, such as rail lines or bridges, is expensive to restructure the temporary change. Junctions, or intersections, are vital components of a road network, all of them regulated by traffic signals. As a result, traffic control signals are a critical component of traffic networks and a primary feature for the successful functioning of traffic networks [3].

Expanding the road network to maximize capacity would be a simple option, but this is not always possible in practice due to space and expenditure constraints. Several strategies have been introduced to overcome the traffic congestion problems, such as speed limit adjustment [4], reinforcement learning for routing [5], neural networks controllers for traffic light [6], hierarchical control method [7], etc. While several variable message sign control algorithms [8] have been developed, most of these methods are reactive in the sense that speed limits are assigned only when congestion is currently observed. Building a real-time traffic management infrastructure that leverages vast volumes of data is a significant problem for current technology, necessitating new technologies to address these obstacles.

Quantum computing is one of the most promising candidates for such technology; first-generation quantum computers and quantum electronics are now commercially accessible in the form of quantum processing units (QPUs), quantum annealers (D-wave systems) [9], ion-trap (IonQ) [10], superconducting quantum computing (Google) [11] and others. If quantum properties such as quantum entanglement and correlations are successfully controlled, it results in quantum algorithms with impressive computational speedups. Quantum annealing has shown potential on combinatorial optimization problems such as job shop scheduling [12], air traffic management [13], trading trajectory problem [14]. Recently, Hussain et. al. [15] have demonstrated the potential of quantum annealing on the traffic signal optimization for a simple and symmetric traffic network. There are some other proposals for traffic optimization using quantum annealing, but the majority of studies have concentrated on route optimizations [16 -17].

In this paper, we propose the real-time optimization of traffic signals phase and traffic light time to achieve maximum traffic flow by formulating the problem as quadratic unconstrained binary optimization (QUBO) problem and performing heuristic optimization using simulated quantum annealing. Our study provides the answer to the problem of great interest, "How to tune the traffic lights and traffic light time in such a way that the average traffic flow increases during the rush hour in a city?". Traffic light controls, traffic generation, traffic modelling, and visualization were done using the SUMO [18]. To build the real-time traffic system using simulated quantum annealing, we first implemented the optimization problem for the simple and symmetric intersection containing only cars with artificial traffic condition and later, we explored more complex and large road intersection "Dongda-Keyuan" located in Taichung, Taiwan, which contains mixed traffic flow with dedicated cars and scooter lane, as well as mixed scooter and car lanes. We used the real-time traffic flow conditions for the simulation, the data was provided by the Taichung Science Park (https://www.ctsp.gov.tw/) to make it more realistic to Taiwan's traffic condition for the left turn scooters were only allowed to make a two-stage left turn. We present several different models for traffic light phase change based on QUBO optimization for compliance with current and future technologies. Furthermore, for the optimal traffic flow, we propose dynamic changes in traffic light signal duration using the QUBO optimization.

†Amit Singh is also with Department of Mechanical Engineering, National Yang Ming Chiao Tung University, Hsinchu 300, Taiwan

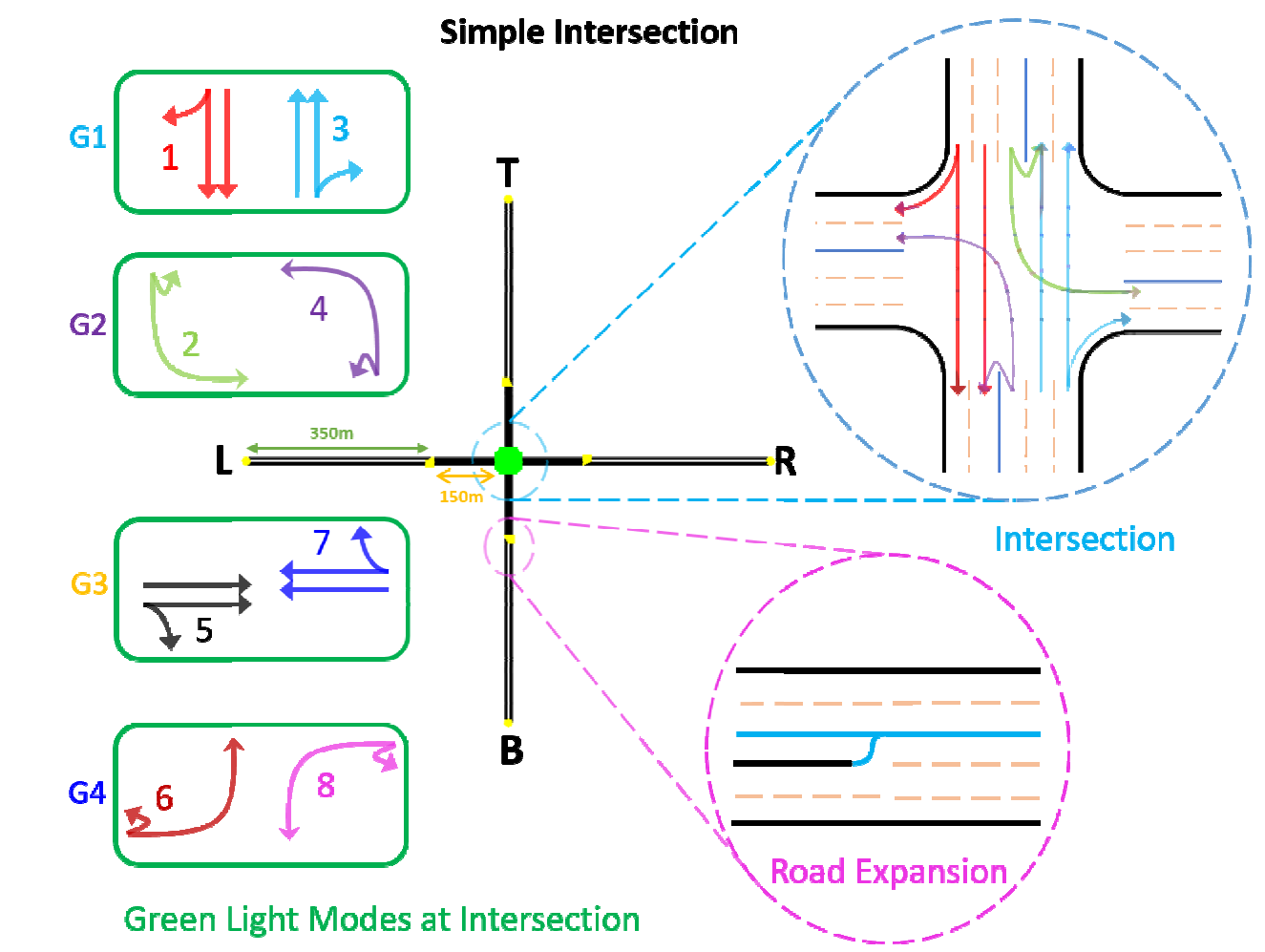

Fig. 1. Schematic of the simple intersection and the traffic light modes at the simple intersection.

This paper is organized as follows. Section 2 introduces the QUBO formulation and optimization technique we used in the study. Section 3 describes the simulation configuration of the intersections and traffic flow conditions used, as well as the real-time traffic considerations in the study. Two QUBO cycles for the traffic signal control were proposed in Section 4. Their results and comparison with usual fixed cycle control is in Section 4. Lastly, we present the conclusions.

## II. QUBO Formulation and Traffic Optimization

Current CMOS annealing machines, digital annealers, and quantum annealers can efficiently solve optimization problems by converting them into an Ising minimization problem, known to be NP-hard on classical computers [9, 19-20]. The Hamiltonian of the Ising model with $n$ binary spins $(\sigma_1, \ldots, \sigma_n)$, external magnetic field $(h_i)$ and interaction between spins $(J_{ij})$, defined as

$$H(\sigma_1, \ldots, \sigma_n) = -\sum_{i<j} J_{ij}\sigma_i\sigma_j - \sum_i h_i\sigma_i \quad (1)$$

is equivalent to the objective function in the QUBO problem,

$$O \coloneqq x^T Q x \quad (2)$$

where $x$ is a vector of $n$ binary variables and $Q$ is the $n \times n$ matrix describing their correlation in between. We used PyQUBO [21] to solve the QUBO problem, which uses *dimod* sampler developed by the D-Wave systems [22]. The developed code can be easily integrated into other Ising machines such as digital annealers, quantum annealers, etc.

In the traffic signal control, the objective function to be minimized in the QUBO problem can be expressed as [15]:

$$O = \sum_j C_{ij} q_{ij} + \varphi \sum_i \left(1 - \sum_j q_{ij}\right)^2 \quad (3)$$

where $q_{ij}$ are binary variables representing whether the traffic light signal mode $j$ (green, yellow, red phase) at the $i$-th junction (traffic light group) is on. Provided with properly chosen coefficient $C_{ij}$, the first term in the right hand side would be associated with the potential the current set of signal modes can improve the overall traffic condition. In this work, we chose $C_{ij}$ as the total number of halting cars on inbound lanes controlled by signal mode $j$ of the $i$-th junction. The objective function by design would optimize the system to clear out the maximum number of vehicles halting before the intersection. The second term with a large enough penalty factor $\varphi$ is the constraint that ensures only one mode would be selected at a time at each junction.

## III. Simulation Setup: Intersection and Traffic Flows

As traffic intersections are the primitive element in a traffic network, we model two different intersection model in the SUMO simulator to study the traffic signal optimization using QUBO: a simple symmetric intersection and the "Dongda-Keyuan" intersection with more asymmetric traffic flows and complicated connection patterns located at the Taichung Science Park of Taiwan (24°12'14.3"N 120°36'38.9"E). Both contain four green light modes as shown in Fig. 1 and Fig. 2. The geometry and the traffic flows used in the simulation for each intersection are described in detail as following:

### A. Simple Intersection

As illustrated in Fig. 1, the crossroad involves three inflow and two outflow lanes at each side (L, R, T, and B), including a dedicated, car-only left-turn (or U-turn) lane extended laterally from the main road 150 meters before the junction. The middle inflow lane is for thru traffic; however, in the real intersections it is quite common for a scooters to sit in this lane with cars under heavy traffic, we thereby relax the restriction and deliberately allow cars and scooters to coexist in the middle inflow lane in the Dongda-Keyuan intersection. The outermost lane is for thru/right-turn traffic for both types of vehicles. This type of junction appears commonly in many large intersections in Taiwan.

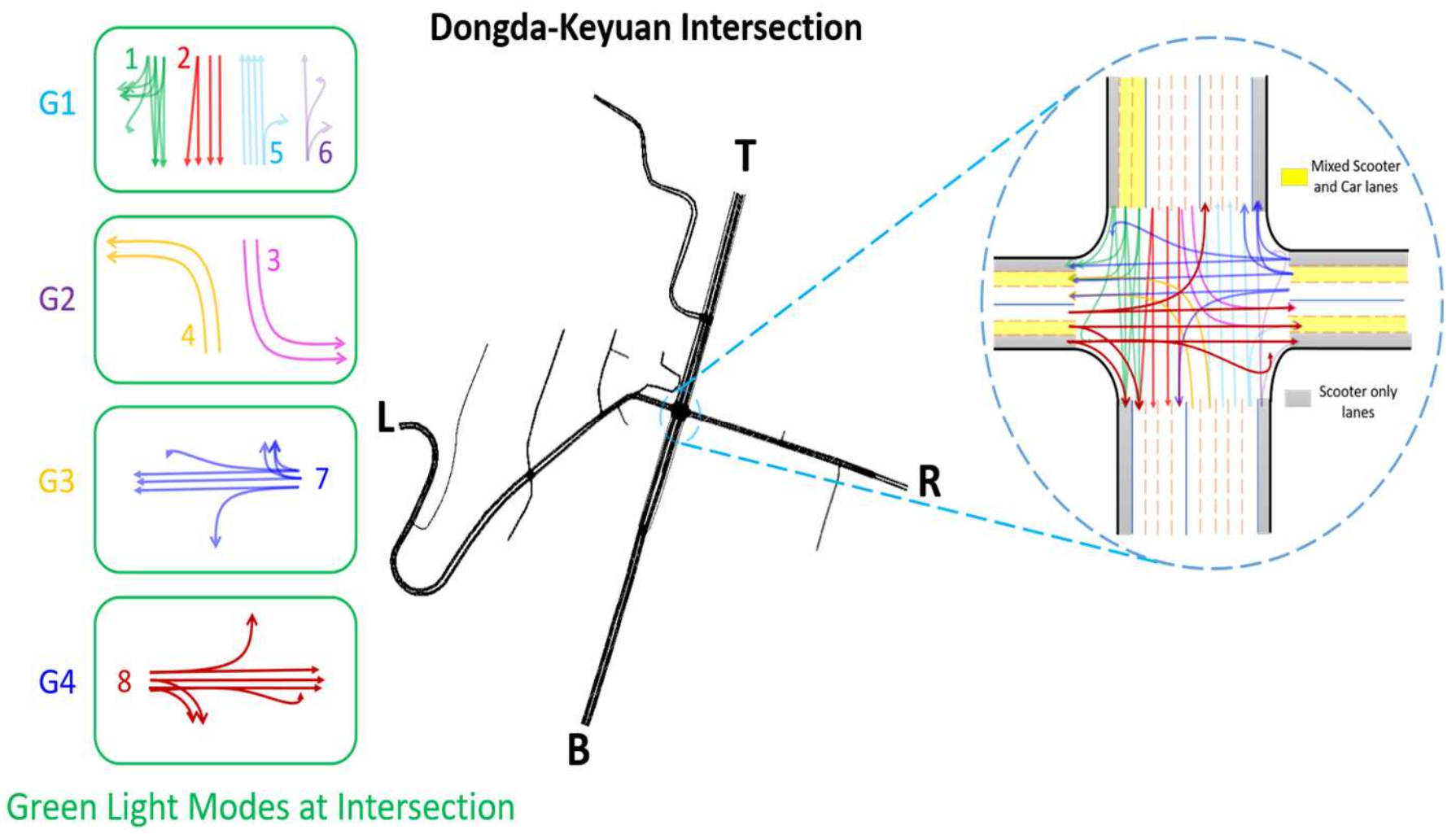

Fig. 2. Schematic of the Dongda-Keyuan intersection (24°12'14.3"N 120°36'38.9"E) at Taichung, Taiwan.

For simplicity, a car-only traffic flow was considered here, leaving the simulation including scooter to the next, more complex intersection. We generated a number of artificial car-only traffic flows, starting from one side of the network to the other, to mimic six types of dense traffic conditions for the optimization study of congested urban traffic. As shown in Table. 1, they are highly dense from all directions (H-Dense), dense from all directions (Dense), dense from the left side (L-Dense), dense from the left and top side (LT-Dense), dense from the left and right sides (LR-Dense) and dense from the left, right, and bottom sides (LRB-Dense). The number of vehicles is loaded uniformly depending on the given traffic conditions (Fig. 3a).

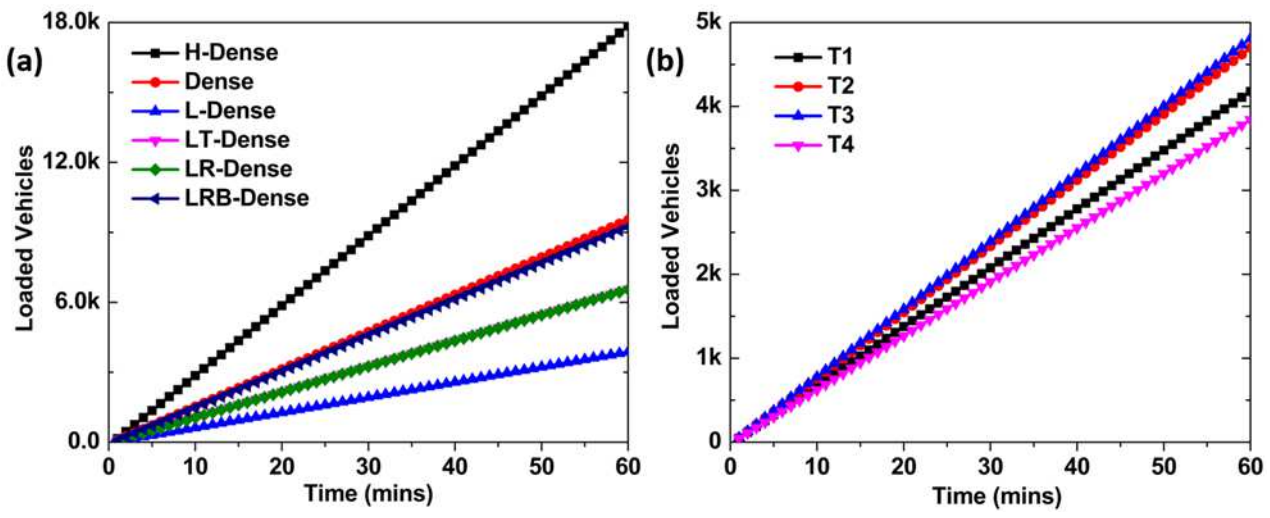

Fig. 3. The average number of vehicles loaded in the traffic network for different traffic conditions in (a) simple intersection and (b) Dongda-Keyuan intersection.

### B. *Dongda-Keyuan Intersection*

In the attempt to apply the optimization study in a real scenario, we choose the "Dongda-Keyuan" road intersection, a heavy-traffic hot spot at the Taichung Science Park, as our target model, as depicted in Fig. 2. We imported the nearby road network from OpenStreetMap via SUMO's OSM tools and carefully configured the intersection with realistic connection patterns followed by cars and scooters that only the indirect, two-stage left-turn is allowed. The top side (T) of the intersection consists of seven inflow lanes and five outflow lanes. These inflow lanes include four car-only lanes, one scooter lane and two mixed lanes (for both cars and scooter); the car-only lanes include a dedicated left-turn lane, a lane for thru/left-turn traffic, two through lanes, and a lane for thru/right-turn traffic. As both the scooter lane and the mixed lane allows thru/right-turn traffic, it could happen sometimes, both in simulation and in reality, that right-turn vehicles at the inner side have to wait for scooters at the outer side. The outflow lanes include four car-only lanes and one scooter lane. The bottom side (B) of the intersection consists of six inflow lanes and five outflow lanes. The inflow lanes include five car-only lanes and a scooter lane, where the former include a dedicated left-turn lane, a lane for thru/left-turn traffic, two thru lanes, and a thru/right-turn lane, and the latter allows both thru/right-turn traffic. The outflow lanes include four car-only lanes and one scooter lane.

TABLE I. SIMULATED TRAFFIC CONDITIONS, WHERE L, R, T, AND B REPRESENT LEFT, RIGHT, TOP AND BOTTOM, LR SHOWS THE NUMBER OF CARS GOING FROM LEFT TO RIGHT.

| **Cases** | **Possible Traffic Congestion Conditions** | | | | | | | | | | | |
|---|---|---|---|---|---|---|---|---|---|---|---|---|
| | ***LR*** | ***LT*** | ***LB*** | ***TR*** | ***TB*** | ***TL*** | ***RB*** | ***RL*** | ***RT*** | ***BL*** | ***BT*** | ***BR*** |
| Highly-Dense | 1500 | 1500 | 1500 | 1500 | 1500 | 1500 | 1500 | 1500 | 1500 | 1500 | 1500 | 1500 |
| Dense | 800 | 800 | 800 | 800 | 800 | 800 | 800 | 800 | 800 | 800 | 800 | 800 |
| L Dense | 1000 | 1000 | 1000 | 100 | 100 | 100 | 100 | 100 | 100 | 100 | 100 | 100 |
| LT Dense | 1000 | 1000 | 1000 | 1000 | 1000 | 1000 | 100 | 100 | 100 | 100 | 100 | 100 |
| LR Dense | 1000 | 1000 | 1000 | 100 | 100 | 100 | 1000 | 1000 | 1000 | 100 | 100 | 100 |
| LRB Dense | 1000 | 1000 | 1000 | 100 | 100 | 100 | 1000 | 1000 | 1000 | 1000 | 1000 | 1000 |

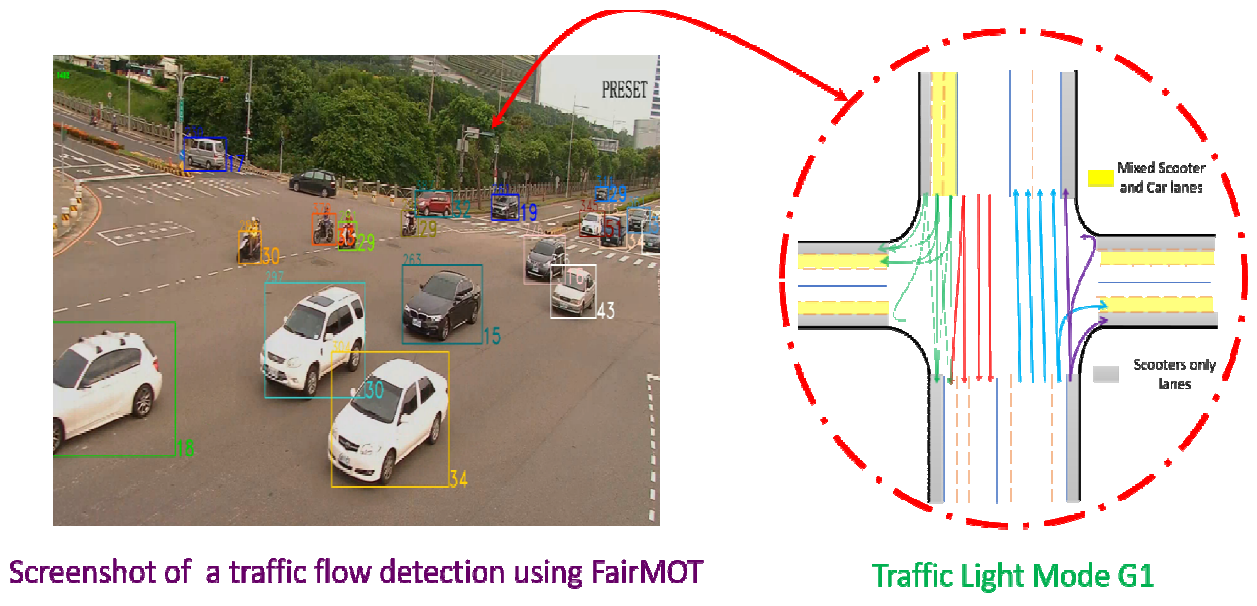

Fig. 4. Snapshot of the traffic flow detection using the cameras installed at the Donda-Keyuan intersection and the number of vehicles and their type was extracted using FairMOT. The boxes in the picture indicate that the vehicle was identified and registered as a unique counting.

The right (R) side of the intersection has the same configuration with the left (L) side, consisting of both three inflow and outflow lanes, whose configurations are very similar to the discussion of the simple intersection in the previous subsection and will be omitted without losing the clarity.

The traffic flow we used in the simulation was inferred from video surveillance of traffic cameras at the intersection during four rush hour period: 7:00-8:00 (T1), 8:00-9:00 (T2), 17:00-18:00(T3), and 18:00-19:00(T4). Based on the number of vehicles registered at each side of the intersection by the FairMOT, a deep learning algorithm for vehicle identification via a single camera developed by Tang. et. al. [23], the amount of traffic flows starting from each side of the network to the others was inferred. Fig. 4 shows a typical snapshot of the vehicle identification using FairMOT when the signal mode G1 was active. With registered vehicle numbers at each side, we estimated 60% of scooters go straight and 20% turn left or right, respectively; we also assumed 30% of vehicles in the right (left) or through lanes make a right (left) turn. To add the variety of vehicle types in the simulation, we assumed 1% of the total vehicles are trucks or trailers. The pattern of overall traffic flow is summarized in Table. 2, indicating the vehicle number injected in the simulation per hour that would start from one side of the network to the other for T1, T2, T3 and T4. For simplicity, the number of vehicles is loaded uniformly from each side as indicated in Fig. 3b.

## IV. Traffic Signal Control

Traffic signals are the primary method for safely and effectively managing scooter and car traffic in a city. Traffic signals must be controlled and calibrated to shift traffic conditions to achieve maximum traffic flow. In traffic management, data analysis is currently used to adjust the fixed cycle timing arrangements for signals [24]. Closely spaced signals are interconnected to create integrated signal networks to optimize traffic volume on long highways. Increasing the average traffic flow speed or reducing the delay time is the key to optimize traffic in a city. In the following subsections, we firstly introduced a traditional fixed-cycle traffic control method, and later, proposed two traffic control cycles using QUBO: the *C-QUBO cycle* and the *QUBO cycle,* both of which are compatible with current and future traffic control methods. In addition, we presented a simple method of dynamically changing traffic light cycle time based on real-time traffic conditions. In future studies, we are planning to employ the reinforced learning or machine learning algorithm to get the optimal real-time traffic signal timing.

### *A. Fixed Cycle*

Fixed-time signal cycle logic in sequential order does not depend on the extra tracking infrastructure, such as pedestrian push buttons or induction loops. Fixed cycle logic is cost-effective but inefficient for the continuously emerging traffic flow. In the simple intersection case, we set 30 secs for the green light followed by 5 secs of yellow light for a smoother transition. While in the Dongda-Keyuan intersection, we adopted the fixed cycles from the real setting provided by the Taichung Science Park: the green light time for mode G1, G2, G3, and G4 are set to be 45, 30, 22, and 24 secs, respectively and 4 secs for the yellow lights (the traffic light modes are shown in Fig. 2). In the fixed-time cycle, signals are in sequential order of G1-G4. Fig. 5a depicts the transition of a fixed cycle signal from a G1 to G3 traffic light.

TABLE II. Traffic flow data of cars and scooter at the Dongda-Keyuan intersection, the T1, T2, T3 and T4 represent the traffic flow of four different rush hour time of 7:00-8:00, 8:00-9:00, 17:00-18:00 and 18:00-19:00 respectively. The L, R, T, and B represent left, right, top and bottom, the combination of directions show the traffic flow directions.

| **Traffic Flow** | **T1** | | **T2** | | **T3** | | **T4** | |
|---|---|---|---|---|---|---|---|---|
| | **Cars** | **Scooter** | **Cars** | **Scooter** | **Cars** | **Scooter** | **Cars** | **Scooter** |
| **LR** | 351 | 311 | 399 | 371 | 434 | 375 | 288 | 288 |
| **LT** | 134 | 104 | 142 | 124 | 153 | 125 | 117 | 96 |
| **LB** | 97 | 104 | 104 | 124 | 115 | 125 | 115 | 96 |
| **TR** | 184 | 79 | 188 | 86 | 148 | 96 | 171 | 64 |
| **TB** | 236 | 238 | 269 | 257 | 220 | 288 | 190 | 192 |
| **TL** | 113 | 79 | 141 | 86 | 142 | 96 | 116 | 64 |
| **RB** | 120 | 46 | 115 | 57 | 72 | 58 | 68 | 47 |
| **RL** | 151 | 226 | 188 | 284 | 187 | 285 | 150 | 232 |
| **RT** | 80 | 46 | 76 | 57 | 48 | 58 | 46 | 47 |
| **BL** | 224 | 131 | 237 | 150 | 256 | 161 | 195 | 130 |
| **BT** | 448 | 394 | 475 | 448 | 512 | 484 | 389 | 388 |
| **BR** | 176 | 131 | 199 | 150 | 217 | 161 | 144 | 130 |

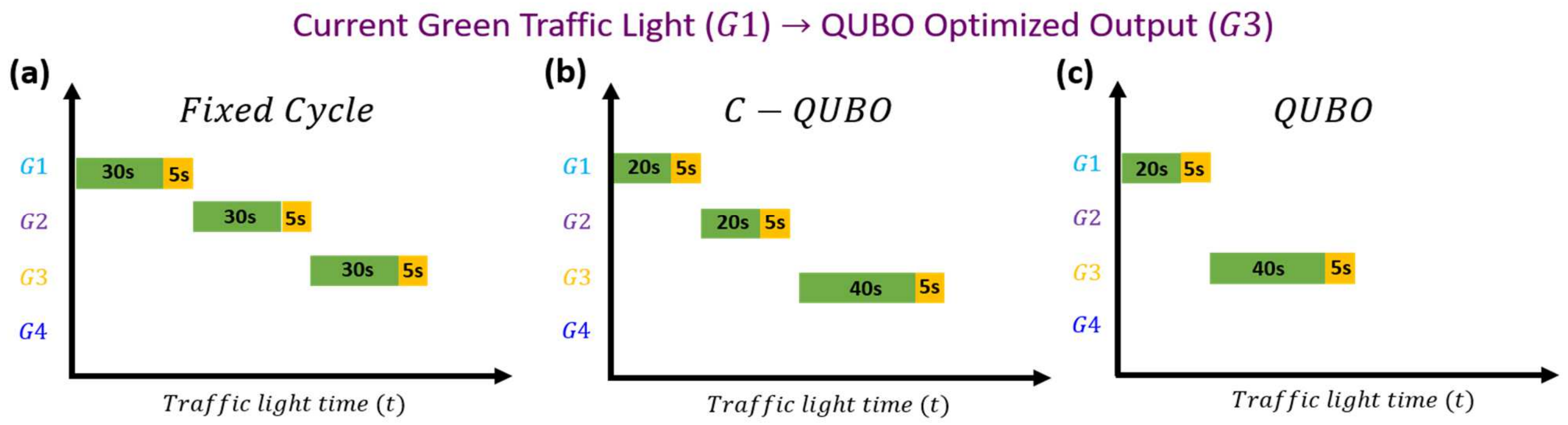

Fig. 5. Traditional and proposed traffic signal cycles methods: (a) Fixed Cycles, (b) C-QUBO cycles and (c) QUBO Cycles.

### *B. C-QUBO Cycle*

In the "cycled"-QUBO (*C-QUBO)* method, the signals cycle in the same manner as the fixed cycle, following the predetermined sequential from G1 to G4. The optimizer proposed a suggested signal based on the simulated traffic flow every 10 seconds. As long as the proposal differs from the current phase, the signal control will be triggered to advance to the next phase after the minimal duration of the green phase and yellow transition phase. For example, if C-QUBO proposes G3 and the actual one is G1, we set 20 secs for the passing green light phase G1 and G2, 40 secs for the proposed G3, and 5 secs for the yellow light phase for the smooth (Fig. 5b). To get the maximum performance, the C-QUBO is also performed once the traffic signal switches modes. The ad hoc duration of (20) 40 secs for the (non-)optimal signals could be further generalized through approaches such as machine learning.

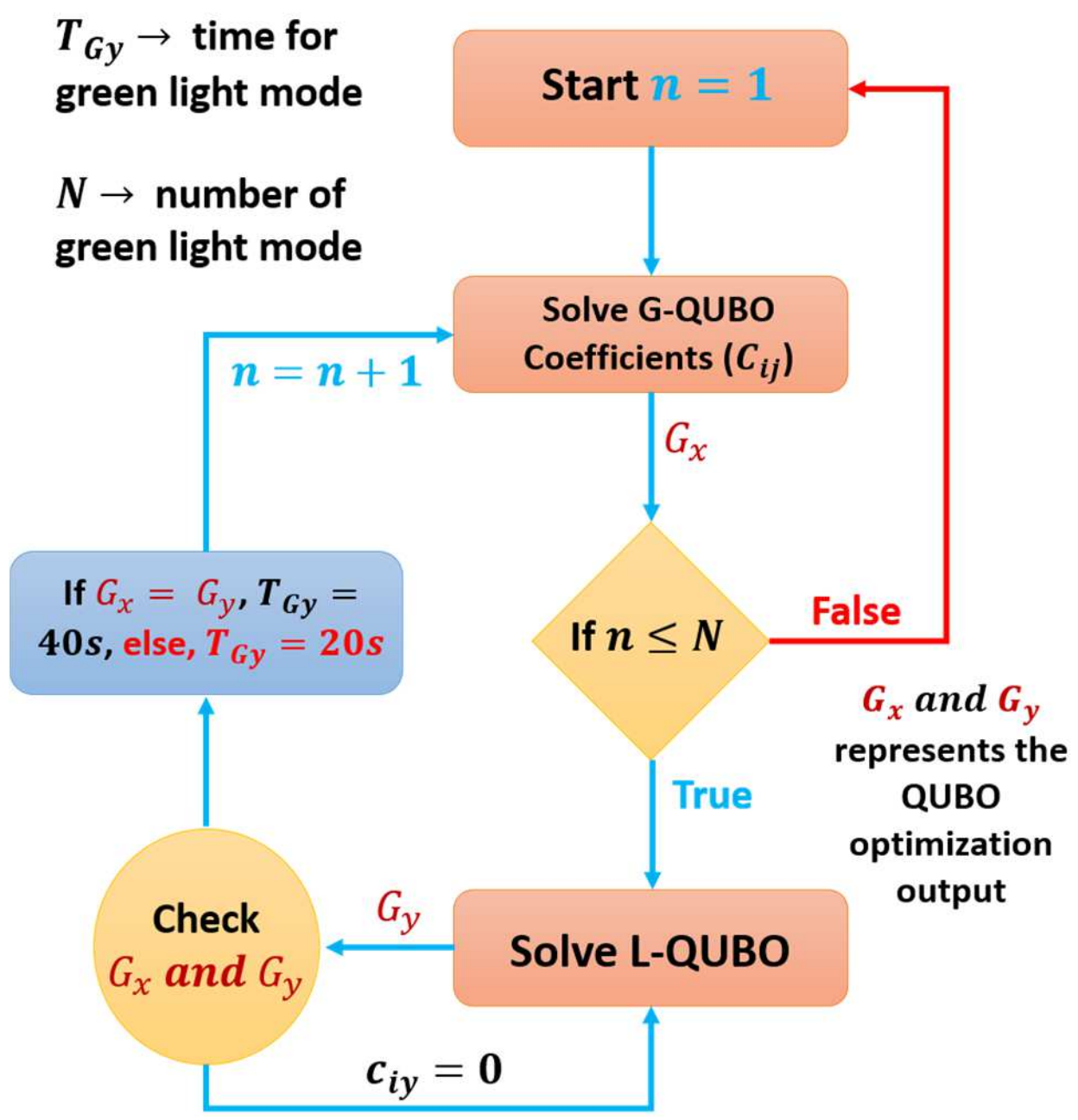

Fig. 6. The logic flow diagram of the two-stage QUBO for fair sharing and to ensure each green light is selected within a phase cycle.

### *C. QUBO Cycle*

In the more aggressive QUBO logic, we allow the optimized traffic signal to hop from one phase to another in a non-predetermined order with a minimum green light phase of 20 secs and yellow transition phase of 5 secs for a smooth transition, as shown in Fig.5c. Our proposed QUBO cycle with a fair-sharing mechanism is shown in Fig.6. To avoid a certain phase to be kept selected by the naive QUBO, which may easily occur in a scenario with a highly unbalanced flow demand, we design the two-stage QUBO to ensure each green phase is selected within a phase cycle group (that comprises N green phases). As the traffic signal initiates a new cycle group, starting with $n = 1$, we run a "global"-QUBO (G-QUBO) for $G_x$, with the coefficients $C_{ij}$, and a "local"-QUBO (L-QUBO) for $G_y$, with a temporary $c_{ij} = C_{ij}$ and after which set $c_{iy} = 0$ to keep $y$ mode from being selected in the next L-QUBO within the cycle group. If $G_x = G_y$, we let the green phase last for 40s; otherwise, we keep it the minimal 20s for fast-rolling into the next preferred green phase.

## V. Results and Discussion

In the traffic simulation via SUMO simulator on both a simple, symmetric intersection and the more realistic and complex Dongda-Keyuan intersection, we found the proposed C-QUBO and QUBO cycles logic delivered a higher average speed per hour of overall vehicle flow over the traditional fixed cycle logic.

To be specific, in the case of a simple intersection under six types of dense traffic conditions as listed in Table 1, we observed that the C-QUBO and QUBO cycle show a significant improvement on the average speed for cases of L-Dense, LR-Dense, and LRB-Dense, while the improvement is marginal for the rest, more symmetric cases (Fig.7a). We saw clearly the relative improvement over the fixed cycle scheme in Fig. 7b. The C-QUBO outperforms the fixed cycle in all except the LT-Dense case, which may be attributed to the randomness of the vehicle behaviors in the simulator, while the QUBO cycle always outperforms the fixed cycle (Fig.7b). The C-QUBO improves average speed by 1.90% (H), 1.16% (D), 18.35% (L), 23.65% (LR), and 8.72% (LRB). While the QUBO improves by 1.21% (H), 0.72% (D), 19.95% (L), 1.36% (LT), 27.00% (LR) and 8.68 % (LRB).

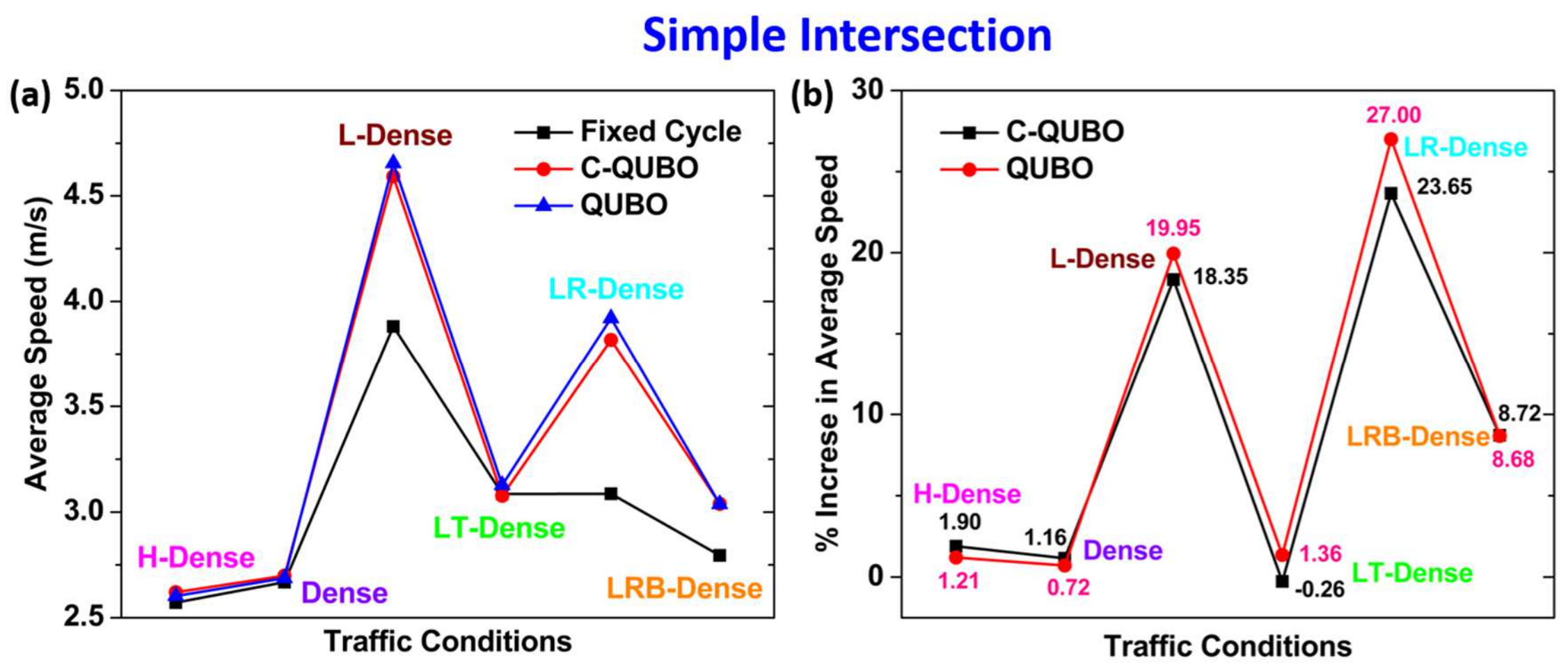

Fig. 7. (a) The average flow speed per hour of the fixed cycle, C-QUBO, and QUBO; (b) the difference in average flow speed per hour of C-QUBO and QUBO relative to the fixed cycle method under various traffic conditions for the simple intersection.

Following that, we investigated the "Dongda-Keyuan" intersection with four traffic flow conditions T1, T2, T3, and T4 that inferred from real-time data as described in Sec. 3. It is found that both the C-QUBO and QUBO cycle performs better than the fixed cycle method. From the average velocity improvement for different traffic conditions (Fig. 8a). In T1, T2, T3, and T4 traffic conditions, the C-QUBO increases average speed by 10.5%, 14.9%, 12.0%, and 10.8%, respectively (Fig. 8b), while the QUBO improves average flow speed by 35.2%, 28.2%, 18.2%, and 28.9%, respectively (Fig. 8b). It is evident from the above that the optimum selection of traffic signal cycle and time will help to reduce traffic congestion.

To further improve the QUBO-based schemes, we proposed a simple way to dynamically change the traffic signal time, by assigning the green light mode duration ($T_{Gy}$) proportional to the number of cars halting for the specific green light :

$$T_{Gy} = C_{ij} \times t_d \tag{4}$$

where $C_{ij}$ is the number of cars halting before the junction $i$ due to the inactive of the the traffic light mode $j$ associated to the $Gy$ mode, and $t_d$ is the average time needed for one vehicle to pass the junction.

We set a minimum duration for a green light mode ($T_{Gy}$) to be 10 seconds to ensure a smooth transition. To get the optimal traffic flow for the Dongda-Keyuan intersection using the dynamic traffic light time, we checked the average flow speed by varying $t_d$ ranging from 0.1 to 1 for the C-QUBO and QUBO cycles (Fig. 9).

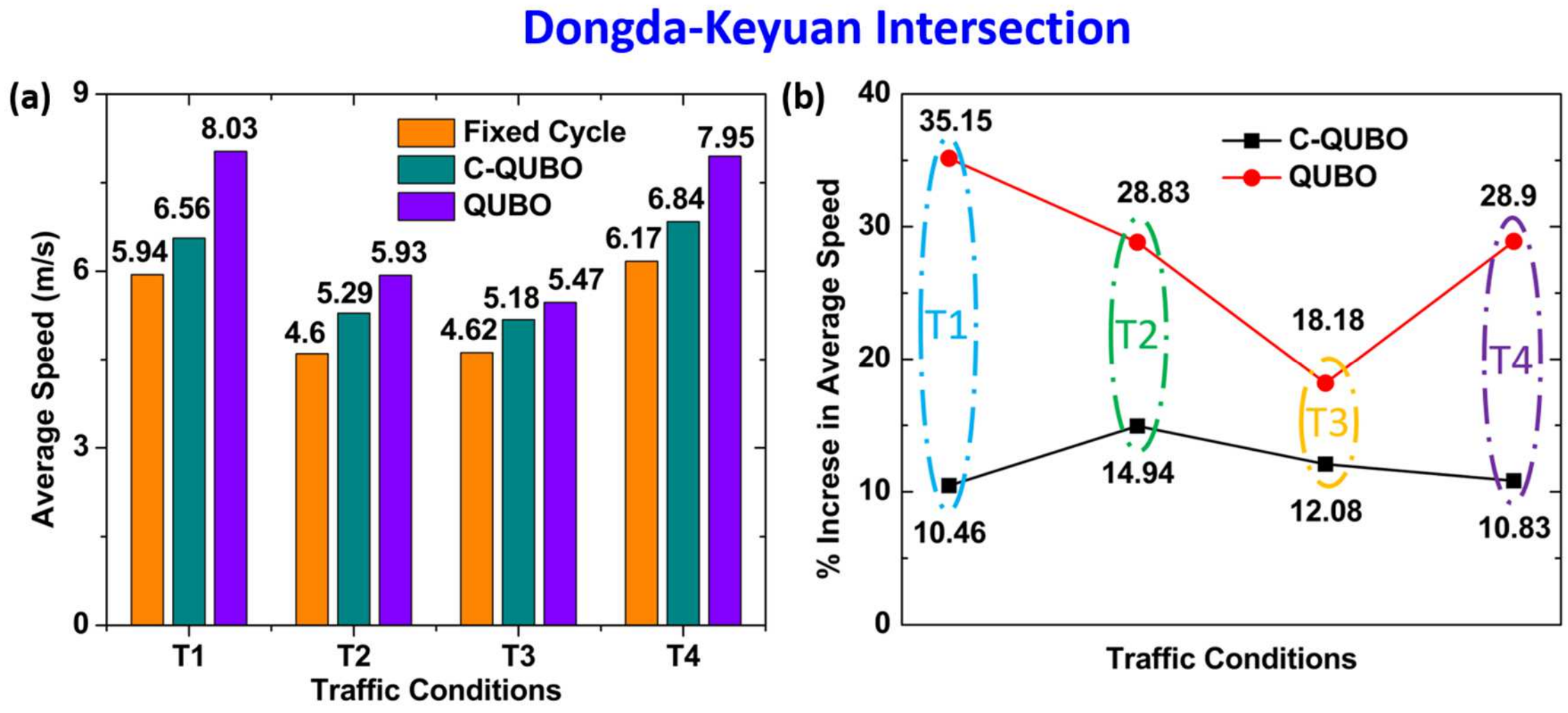

Fig. 8. (a) The average flow speed per hour of the fixed cycle, C-QUBO, and QUBO; (b) the difference in average flow speed per hour of C-QUBO and QUBO relative to the fixed cycle method under various traffic conditions for the Donda-Keyuan intersection.

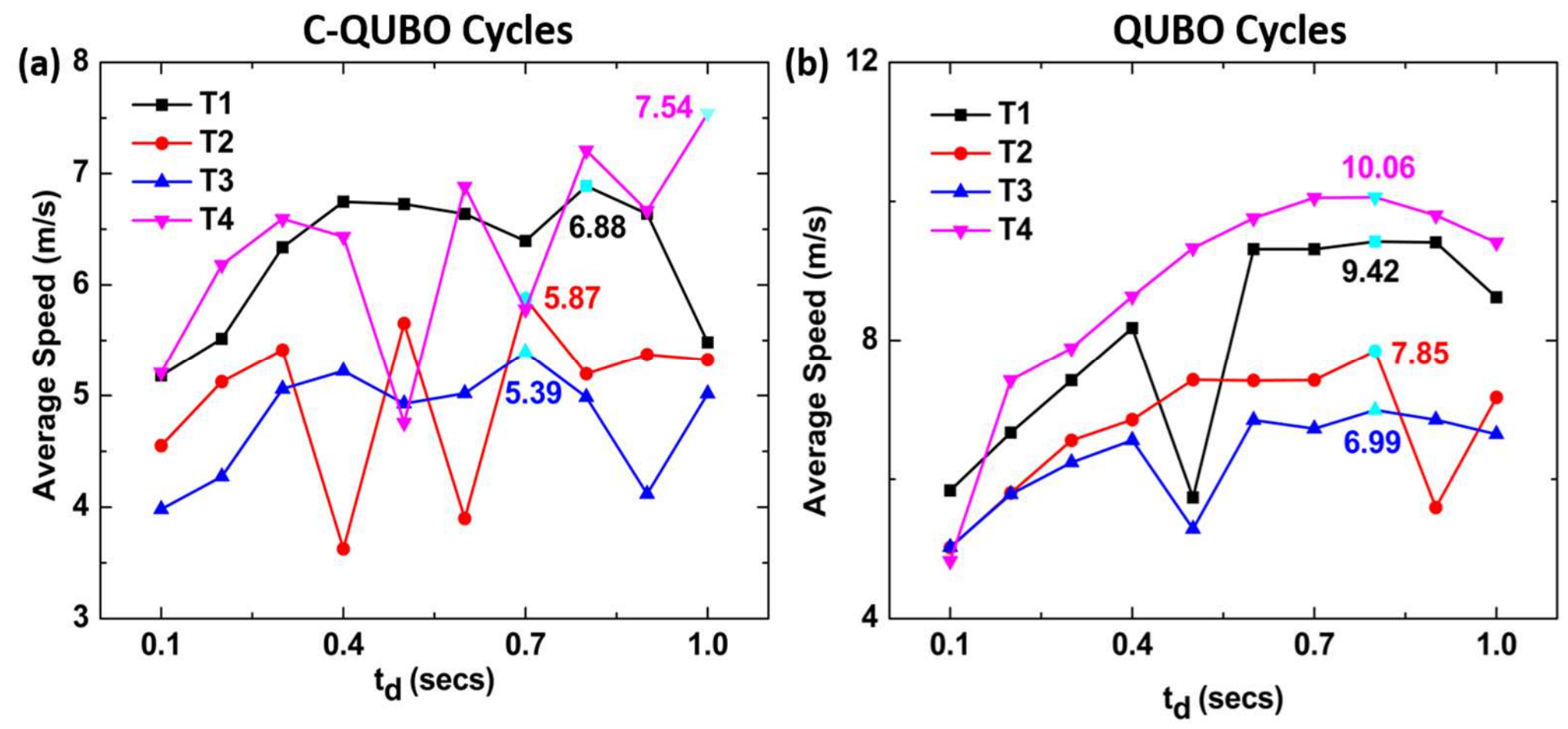

Fig. 9. The average flow speed per hour of the (a) C-QUBO and (b) QUBO methods with dynamical green light mode for the Donda-Keyuan intersection under different traffic conditions. The maximum average flow speed for the various traffic conditions are represented by the cyan color marker.

The maximum average flow speed in the C-QUBO cycles is found at $t_d$ values of 0.8, 0.7, 0.7 and 1 for the traffic conditions T1, T2, T3 and T4, respectively, while in the QUBO cycles it is found at $t_d$ value of 0.7 for all the traffic conditions T1, T2, T3 and T4 (Fig. 9).

Next, we choose the most efficient average flow speed to compare the C-QUBO and QUBO with optimal $T_{Gy}$ relative to the fixed cycle method for the Dongda-Keyuan intersection. In T1, T2, T3, and T4 traffic conditions, the C-QUBO with optimal $T_{Gy}$ increases average speed relative to the fixed cycle method by 15.90%, 27.60%,16.59%, and 22.23%, respectively (Fig. 10). While the QUBO improves average flow speed by 58.55%, 70.45%, 51.11%, and 63.06% in T1, T2, T3, and T4 respectively (Fig. 10).

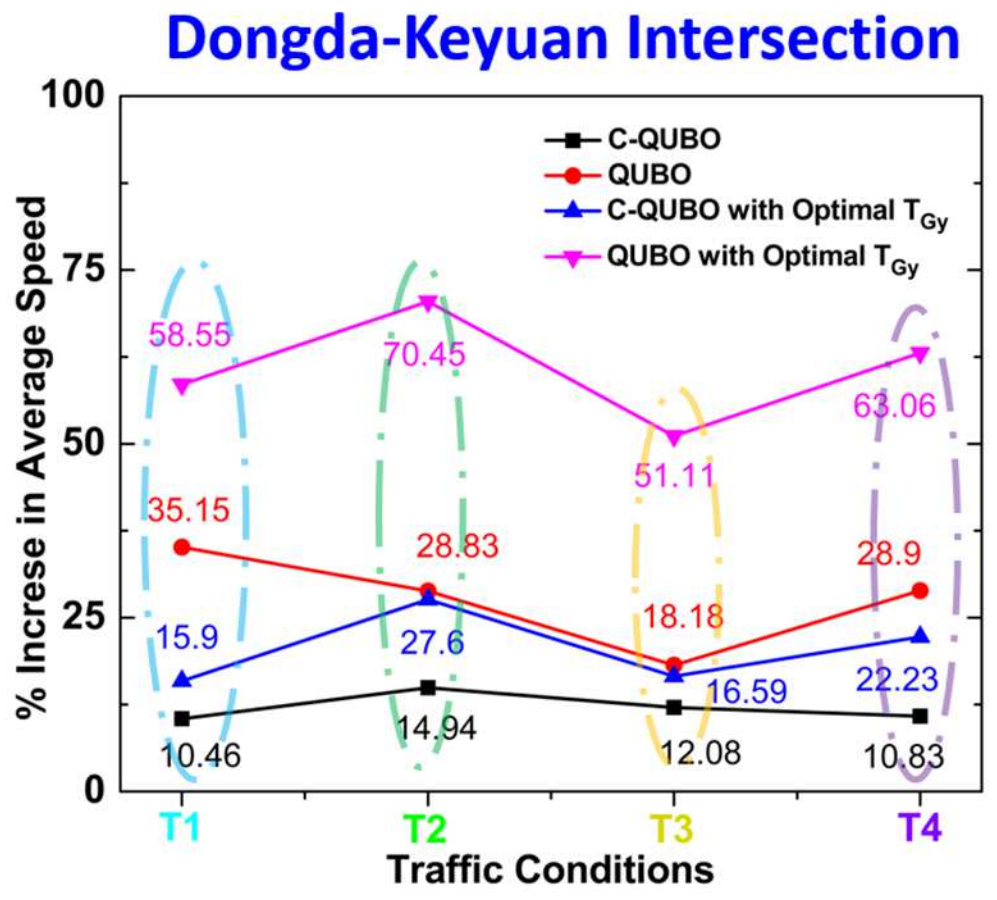

Fig. 10. The difference in average flow speed per hour of C-QUBO, QUBO, C-QUBO with optimal $T_{Gy}$ and QUBO with optimal $T_{Gy}$ relative to the fixed cycle method for the Dongda-Keyuan intersection under various traffic conditions .

## VI. Conclusion

In conclusion, we proposed a quantum annealing approach to effectively maximize the traffic flows at an intersection through optimal traffic signal control. The ultimate goal is to relieve congestion, pollution, travel delays, and increased travel expenses incurred by traffic congestion in urban areas. It is found that in the simple intersection case, when the flow is concentrated in the left and right directions (LR-Dense), C-QUBO and QUBO demonstrate the maximum increases in the average speed per hour of total flow by 23.6% and 27.0 %, respectively. In the real intersection "Dongda-Keyuan" scenario, the C-QUBO and QUBO show maximum improvements of 14.9 and 35.1% in T2 and T1 traffic condition cases, respectively. Furthermore, it is found that dynamically adjusting the traffic light time will provide us further increase in the average speed of traffic flow. The dynamical change in the traffic signal duration and selection of the optimal traffic light time for the Dongda-Keyuan intersection results in a maximum improvement in average velocity per hour by 35.15% and 70.45% relative to fixed cycles for the T1 and T2 traffic conditions, respectively.

The current approach's significant improvement in traffic flows will lead us to investigate more complicated and interlocked road configurations, such as grids. Our ultimate goal is to optimize traffic flow on a city-wide scale, expanding QUBO optimizations can be done by considering the correlation of traffic lights by including cost terms with nearest neighbor correlation. QUBO can also be applied directly to quantum annealing machines, such as D-WAVE, allowing us to perform calculations on a larger scale in the near future.

## Acknowledgment

This paper was written as part of the research internship at the National Center for High-Performance Computing and National Applied Research Laboratories, Hsinchu, Taiwan. We are immensely grateful to Central Taiwan Science Park (CTSP) for the field support. We are also grateful to Dr. Jenq-Neng Hwang and his group for the help on traffic tracking and to Dr. Lien-Po Yu on useful discussions on quantum computing.